\documentclass[%
 preprint, 
superscriptaddress,
 amsmath,amssymb,
 aps, physrev,
]{revtex4-2}

\usepackage{graphicx}
\usepackage{dcolumn}
\usepackage{bm}
\usepackage[utf8]{inputenc}
\DeclareUnicodeCharacter{2212}{\textminus}

\usepackage{xcolor}
\begin{document}


\title{\textbf{In-vivo femtonewton-sensing nanotribology of \textit{Tradescantia zebrina} leaf cell inner surface using roll rotation detection} 
}%

\author{Snigdhadev Chakraborty}
\author{Mukul Sagar}%
 
\affiliation{%
 Department of Physics, Quantum Centre of Excellence for Diamond and Emergent Materials (QuCenDiEM), IIT Madras, Chennai 600036, India
}%

 

 

\author{Atanu Ghosh}%
 
\affiliation{%
 Department of Physics, Quantum Centre of Excellence for Diamond and Emergent Materials (QuCenDiEM), IIT Madras, Chennai 600036, India
}%

\author{Krishna Kumari Swain}
\affiliation{%
 Department of Physics, Quantum Centre of Excellence for Diamond and Emergent Materials (QuCenDiEM), IIT Madras, Chennai 600036, India
}%

\author{Mrutyunjaya Rath}
\affiliation{%
 Department of Physics, Quantum Centre of Excellence for Diamond and Emergent Materials (QuCenDiEM), IIT Madras, Chennai 600036, India
}%

\author{Agniva Das}
\affiliation{%
 Department of Physics, Quantum Centre of Excellence for Diamond and Emergent Materials (QuCenDiEM), IIT Madras, Chennai 600036, India
}%

\author{Susy Varughese}
\affiliation{%
 Department of Chemical Engineering, IIT Madras, Chennai 600036, India
}%

\author{Basudev Roy}

\email{Contact author: basudev@iitm.ac.in}

\affiliation{%
 Department of Physics, Quantum Centre of Excellence for Diamond and Emergent Materials (QuCenDiEM), IIT Madras, Chennai 600036, India
}%


\begin{abstract}

Accessing the properties of a plant cell interior non-invasively is difficult due to the presence of a cell wall. Nanoparticles larger than 5 nm cannot be readily phagocytosed inside the cell like animal cells. It is here that we realise that \textit{Tradescantia zebrina} plant cells have prismatic forms of calcium oxalate crystals present inside them naturally. These crystals make a ready choice to study properties of the inner cell surface with the application of optical tweezers. Moreover, out-of-plane rotations in optical tweezers have begun to be explored only recently. The pitch rotation has been detected with high resolution and several applications are explored. In this work, we first study the stable configuration while trapped in linearly polarized optical tweezers and then explore the other out-of-plane configurations to detect the roll rotation at high resolution. Then a micro-rheological analysis is performed to obtain the frictional properties of the inner surface of the plasma membrane of the leaf cell. The size of the particle is about 5 $\mu$m along the diagonal, so that the contact length with the surface is about 200 nm. We measure a frictional force of 18.5 pN at a sensitivity of about 200 fN without averaging. 

\end{abstract}

\maketitle


\section{\label{sec:level1}Introduction:}

Plant cells have a rigid cell wall around them, making access to the interior difficult \cite{zhang2024heterogeneity}. Any foreign particle needs to be tens of nanometers or smaller and inserted using complicated techniques, particularly at the time of germination of the plant\cite{wang2023nanoparticles,sembada2024transport}. It is here that native particles inside plant cells attain utility as these can then be used to sense forces and torques, and therefore ascertain the properties of the plant cell interior non-invasively. 

\textit{Tradescantia} group of plant leaves are frequently studied to investigate various aspects of leaf mechanics, including chloroplast characteristics\cite{park1996chloroplast}, turgor pressure\cite{zimmermann1980direct,ye2008cell}, cell wall properties\cite{cleary2001plasma}, and photosynthesis efficiency in response to light and CO$_2$ \cite{anderson2001response}. Within these leaves, calcium oxalate (CaC$_2$O$_4$,nH$_2$O) crystals, such as raphides, styloids, and prismatic forms, are present. These crystals are produced through biomineralization and serve as a calcium storage mechanism. In addition, the abundance of calcium oxalate (CaOx) crystals acts as a defence mechanism, deterring herbivores from grazing.
Optical tweezers have emerged as a powerful tool for micro-manipulation with unprecedented accuracy and control. It has revolutionized the field of biophysics\cite{bustamante2021optical,sudhakar2021germanium,betz2009atp}, soft matter physics\cite{paul2018two,robertson2018optical,betz2012time}, nano-photonics\cite{chand2023emergence,paul2022simultaneous,badman2019towards}, and micro-engineering\cite{li2022optical,butaite2019indirect,enger2004optical}.
Researchers have routinely trapped different cytoplasmic organelles like chloroplasts \cite{garab2005alignment,Li2015chloroplasttrap,Gao2015chloroplasttrap}, Golgi bodies \cite{sparkes2009grab}, amyloplasts \cite{abe2020micromanipulation}  inside plant cells for the last two decades. Although a single report\cite{dasgupta2003controlled} has claimed the trapping of intracellular particles presumed to be calcium oxalate, no detailed study has yet demonstrated the optical confinement of such particles within living plant cells. The optical trapping of calcium oxalate crystals is enabled by the refractive-index\cite{wherry1922review} contrast between the crystals and the surrounding cytosolic medium.

 CaOx crystals, primarily as the monohydrate (whewellite) and dihydrate (weddellite) phases, constitute the main mineral component of human kidney stones. These crystals are also detected in urinary sediments \cite{grover1992calcium,werner2021calcium}, where their presence reflects supersaturation of calcium and oxalate ions and indicates an elevated risk of stone‐formation \cite{daudon2016respective}.

Recently, researchers used Atomic Force Microscopy (AFM) \cite{tsugawa2022elastic,pu2022afm} to measure the rheological and physical properties of plant cells and cell walls. However, the presence of the cell wall makes it difficult to probe the contents of the cell from the outside. Microneedles and micropipettes are used as invasive tools to mechanically disrupt the cell wall, allowing access to the intracellular fluid and plasma membrane. In contrast, optical tweezers provide a non-invasive method to trap native organelles, such as chloroplast\cite{vavaev2025laser}, enabling the measurement of their rheological\cite{mukherjee2021active,kundu2021single}and physical properties\cite{vaippully2019study, roy2023comparison1} within the native cellular environment. AFM-based adhesion or membrane tension studies require an initial hard pressing on the surface\cite{jiang2016measurement,sen2005indentation}, which can therefore be referred to as "hard probing." Such hard probing can damage or even kill sensitive biological samples. However, optical tweezers-based probing methods can be used, which require much smaller forces, on the order of 10 pN\cite{peri2008adhesion}. This approach is referred to as  soft probing\cite{friedrich2015surface,miyoshi2010growth}. Because the optical tweezers have a trap stiffness (pN/nm) much lower than that of the AFM, it can measure smaller forces with higher sensitivity.
 Roll rotation is one of the out of plane rotations observed in leukocytes during transport and adhesion to blood vessel wall\cite{mcever2010rolling}. During acute inflammation, leukocytes migrate to the injured tissue by rolling along the endothelial surface of blood vessels\cite{chang2000state,king2005cell}. This rolling arises from a competition between the shear forces of flowing blood and the reversible selectin–ligand bonds that form at the front of the contact patch and dissociate at the rear.

Here, we optically confine one calcium oxalate particle in linearly polarized tweezers and demonstrate that it assumes a side-on configuration with one of its vertices pointed vertically downward. This becomes suitable for probing the properties of the cell membrane. We place the vertex of the crystal on the cell membrane and then move the leaf sample parallel to the surface. This generates out-of-plane rotation of the crystal, which we call roll rotation\cite{nalupurackal2023controlled} in the nomenclature of airlines\cite{roy2018determination}. 

In the case of rolling without slipping, the roll angle($\theta$) is related to the amplitude of the stage motion($S$) as $S=r\theta$. Where r is the radius of the probe particle.
In lateral-force microscopy (LFM), which is widely used to study friction\cite{yang2024atomic,ozdogan2025lateral}, the extent of tip rotation depends on the tip length (typically on the order of 10-100$\mu m$) rather than on the 100 nanometre-scale apex radius. Hence, an optically trapped micrometre-sized probe will undergo a rotation greater than that of an LFM tip, resulting in improved sensitivity. Moreover, the torsional stiffness of the cantilever tips is about 3 orders of magnitude stiffer \cite{cannara2006lateral} than rotational trap stiffness in optical tweezers, making it possible to detect smaller forces with this in the true spirit of soft probing. 
We detect the rotation angle using our novel technique, which relies on the asymmetry of the scatter pattern behind crossed polarizers, allowing for such detection at high resolution and bandwidth. We compare the roll power spectra obtained with the vertex touching the cell membrane and when the crystal is in bulk to extract tribological properties\cite{vaippully2019study,lokesh2021estimation} of the plant cell inner surface. We also demonstrate the successful detection of frictional forces of approximately 1 pN with a sensitivity of around 10 fN. 
 
$$$$$$$$

\section{\label{sec:level2}Theoretical results}

In the context of our simulation setup, the rotation about the z-axis is referred to as yaw, which represents an in-plane rotation. The rotations about the x-axis and y-axis are termed roll and pitch, respectively, corresponding to out-of-plane rotations. The direction of laser propagation is along the z-axis, and the laser polarization is along the y-axis, as shown in FIG. \ref{fig:lumerical_setup3}.

Finite-Difference Time-Domain (FDTD) simulation is a numerical technique that solves Maxwell's equations in complex geometries by discretizing both time and space into a mesh, enabling the computation of electromagnetic field interactions with various materials.

\subsection{\label{sec:level3}  Stability analysis }

\begin{figure}[h!]
\centering\includegraphics[width=.75 \linewidth]{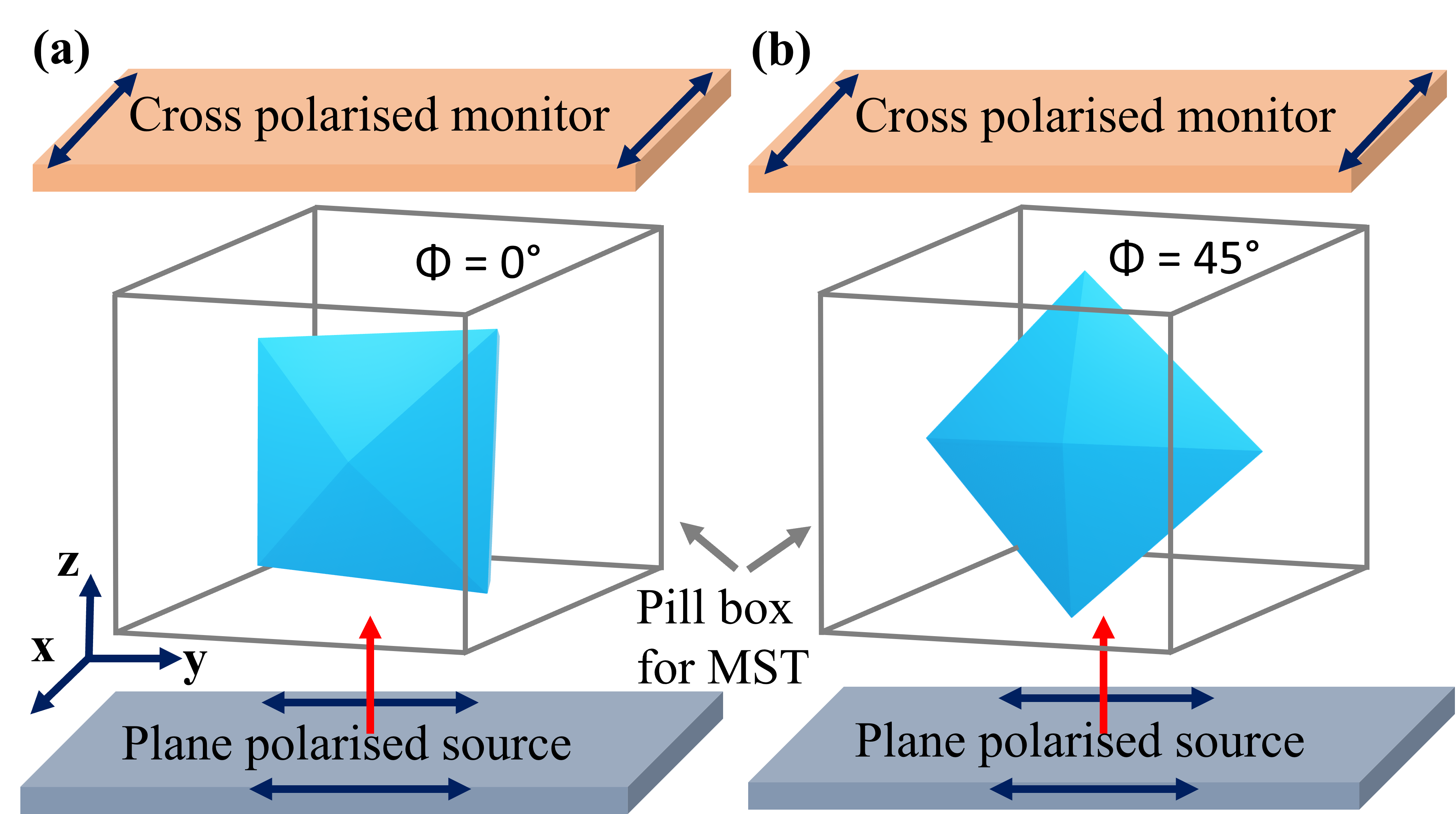}
\caption{Possible configurations of a Calcium Oxalate crystal when trapped in linearly polarized optical tweezers}
\label{fig:lumerical_setup3}
\end{figure}

\begin{figure}[!ht]
\centering\includegraphics[width=1\linewidth]{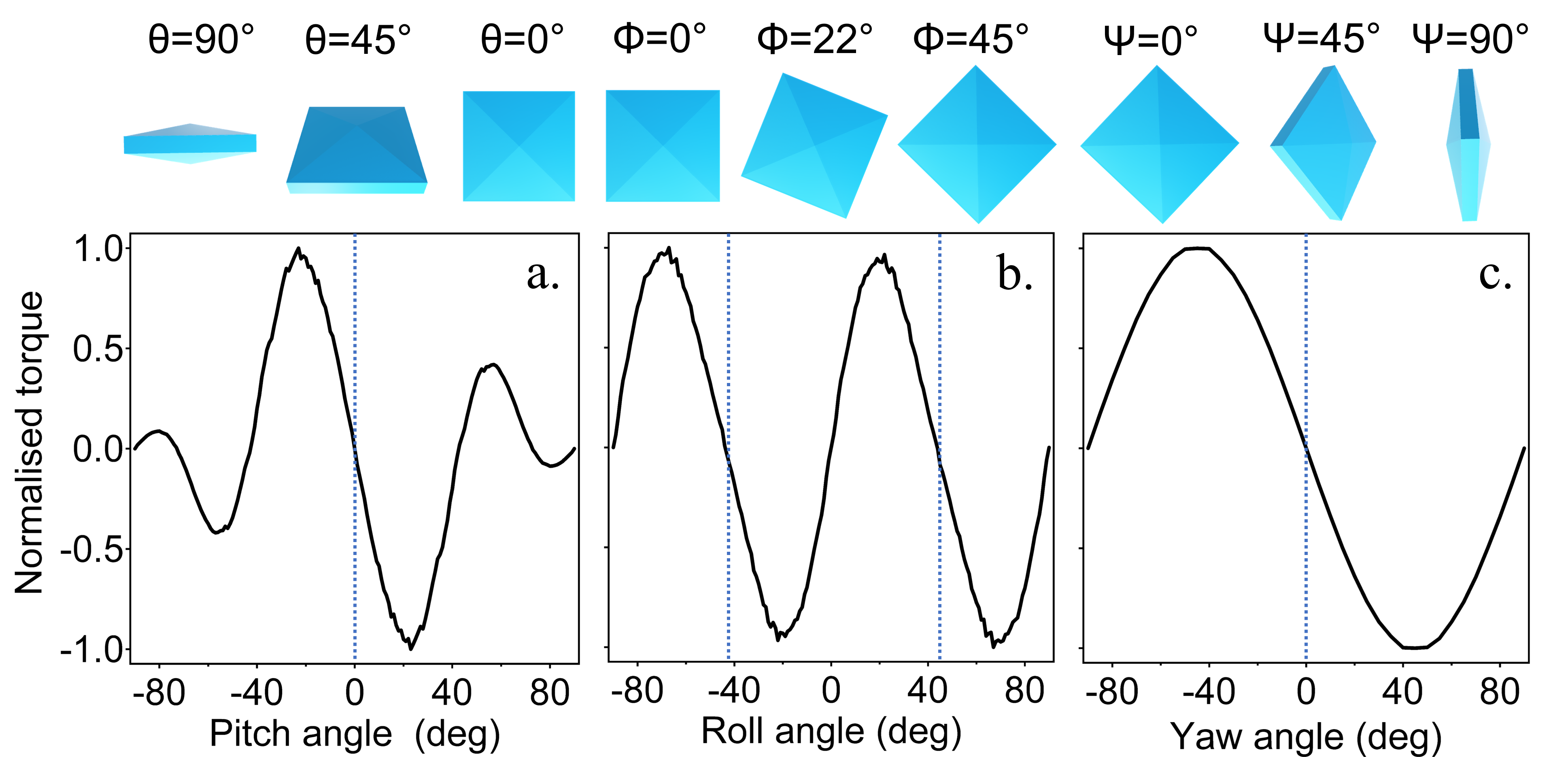}
\caption{Stability analysis of a calcium oxalate crystal based on rotational torque calculations:(a) Torque as a function of pitch rotation; the stable orientation is observed at $\Theta$=0\textdegree.(b) Torque as a function of roll rotation; stable orientations are found at $\Phi$=±45\textdegree(c) Torque as a function of yaw rotation; the stable orientation occurs at $\Psi$=0\textdegree}
\label{fig:torque_analysis}
\end{figure}

Experimental results confirm the presence of axial and radial optical torques acting on the trapped Calcium Oxalate (CaOx) particle. To understand the stable trapping configuration of the trapped particle\cite{cao2012equilibrium,singer2006orientation}, we performed a Lumerical FDTD simulation. A  prismatic-shaped particle of transverse width 4 $\mu$m and thickness 1 $\mu$m was placed inside a square-shaped cubic test region of 10 $\mu$m width. An imaginary bounding box is placed around the particle to compute the Maxwell Stress Tensor (MST) and, consequently, the torque experienced by the particle. We apply a plane-polarized Gaussian laser beam of wavelength 1 $\mu$m into the system as shown in the FIG. \ref{fig:lumerical_setup3}. The optical torques are calculated by integrating the Maxwell stress tensors $T$ over an imaginary surface enclosing the particle. Maxwell stress tensor components are defined as 

\begin{equation}
T_{ij} = \varepsilon E_i E_j + \mu H_i H_j - \frac{1}{2} (\varepsilon E^2 + \mu H^2) \delta_{ij}
\end{equation}
where, $E_i$ and $H_i$ are the components of the electric and magnetic fields, $\varepsilon$ is the permittivity of the medium, $\mu$ is the permeability of the medium, $\delta_{ij}$ is the Kronecker delta. After calculating the stress tensor components, the optical torques are obtained by integrating,
\begin{equation}
\tau_k = \oint_S \epsilon_{klm} r_l T_{mj} dS_j
\end{equation}

where, $\tau_k$ is the $k$-th component of the torque,
$\epsilon_{klm}$ is the Levi-Civita symbol, $r_l$ is the $l$-th component of the position vector relative to the center of the particle, $T_{mj}$ are the components of the Maxwell stress tensor, $dS_j$ is the differential surface element in the $j$-th direction. We first consider the rotation of the CaOx particle about its pitch axis, characterized by the pitch angle $\Theta$. A stable equilibrium configuration is achieved when the optical torque is zero, and its derivative with respect to the rotation angle is negative. We find out that the side-on configuration ($\Theta$=0\textdegree) is the most stable, as shown in FIG. \ref{fig:torque_analysis}a. Then we rotated the side-on oriented particle along the roll fashion, and the stable orientation obtained is the roll angle $\Phi$=±45\textdegree, as shown in FIG. \ref{fig:torque_analysis}b. Here, $\Phi$=0\textdegree  orientation corresponds to the case when two sides of the particle are parallel to the laser polarization, while the other two sides are parallel to the direction of laser propagation. Subsequently, the yaw rotation was performed. The torque-angle graph depicts that the yaw angle $\Psi$=0\textdegree is the most stable orientation. The particles align preferentially with the polarization of the incident laser..

\subsection{\label{sec:level4} Detection method:
}
 Tetragonal bipyramidal-shaped calcium oxalate particles prefer to align side-on when trapped using a linearly polarized laser beam, as discussed in the previous section. We can detect the out-of-plane rotations by exploiting the shape anisotropy of the particle. When a laser is illuminated on this particle, the forward-scattered light passing through a set of crossed polarizers changes with the orientation of the particle. In order to explore the proof of concept, we perform another Lumerical FDTD simulation\cite{lokesh2022generation} to establish the relation between the roll and pitch angle with cross-polarized intensity of the forward scattered pattern. In the simulation, we solve Maxwell's equations with appropriate boundary conditions and geometry.

A  prismatic-shaped particle of transverse width 4$\mu$m and thickness 1$\mu$m was placed inside a square-shaped cubic test region of 10 $\mu$m width. We illuminate the prismatic-shaped particle with plane-polarized light in the simulation and measure the intensity of the orthogonally polarized light using a monitor placed above the particle. We observe that the scattered pattern makes a four-lobe pattern. As the particle rotates in the pitch direction (rotation about the y-axis), an asymmetry between the upper and lower halves is observed. For pitch angle rotation (from −20° to +20°), the difference in intensity between the upper and lower halves varies linearly with the pitch angle. Similarly, for roll rotation (rotation about the x-axis), a noticeable asymmetry appears between the right and left halves. For roll angles in the range of −20° to +20°, the difference in the intensity between the right and left halves is linearly proportional to the roll angle. The results are shown in FIG. \ref{fig:torque_analysis}.

Previous efforts to detect the pitch angle at high resolution have focused on spherical birefringent particles\cite{roy2018determination} and hexagonally shaped upconverting particles\cite{chakraborty2023high}. However, detecting the other out-of-plane rotation, known as roll rotation, remains elusive. In the present work, we consider a prismatic-shaped particle with a geometry that is entirely different from those of the previously studied particles. It is not immediately clear that the cross-polarized signal would naturally apply to this case. Therefore, we performed simulations and found that the difference-in-halves method using crossed polarizers to detect both pitch and roll is also effective in this situation

\begin{figure}[!ht]
\centering\includegraphics[width=1.0 \linewidth]{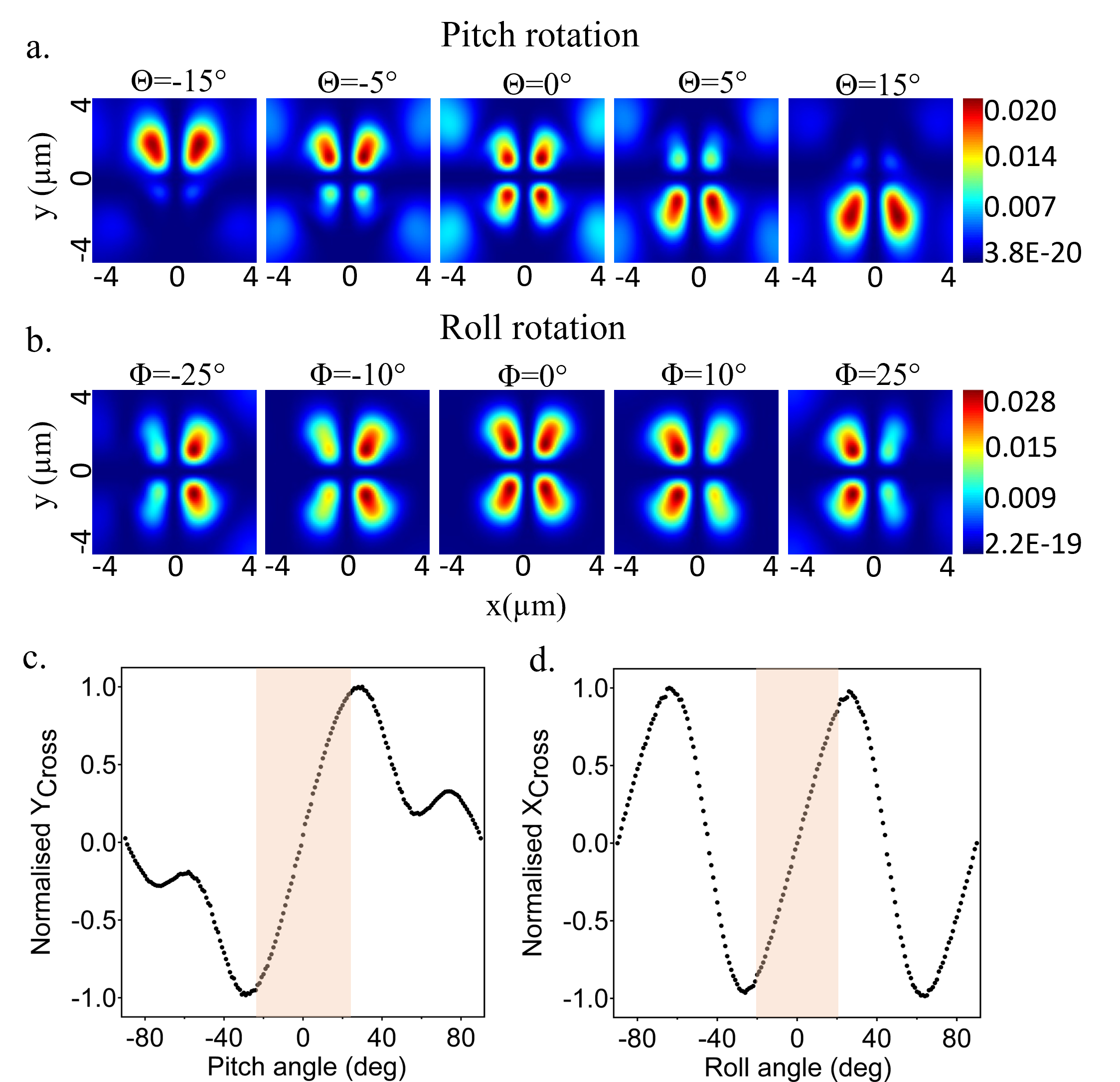}
\caption{This figure shows the roll detection system with photonic force microscopy (a)  The simulated scatter patterns for such a particle exhibiting pitch rotation detected with photonics force microscopy is shown (b) Simulated scatter patterns for such a particle exhibiting roll rotation detected with photonics force microscopy (c) The transfer function for detected signal as a function of pitch angle (d) The transfer function for detected signal as a function of the corresponding roll angle turn}
\label{fig:FOUR LOBE}
\end{figure}

This model in Lumerical is able to predict optical trapping forces by calculating the electric and magnetic components of the fields using a Finite Difference Time Domain (FDTD) solver, which solves the Helmholtz equation for such systems \cite{lokesh2022generation}. This simplistic model computes the effects due to the scattering from the particle. The particle alters the scattered light pattern as a consequence of interaction with the shape and orientation of the particle. The effect is indicated by the FDTD simulations. The results are shown in FIG. \ref{fig:FOUR LOBE}.

\section{\label{sec:level5}Experimental results and discussions}

\begin{figure}[!ht]
\centering\includegraphics[width=.8 \linewidth]{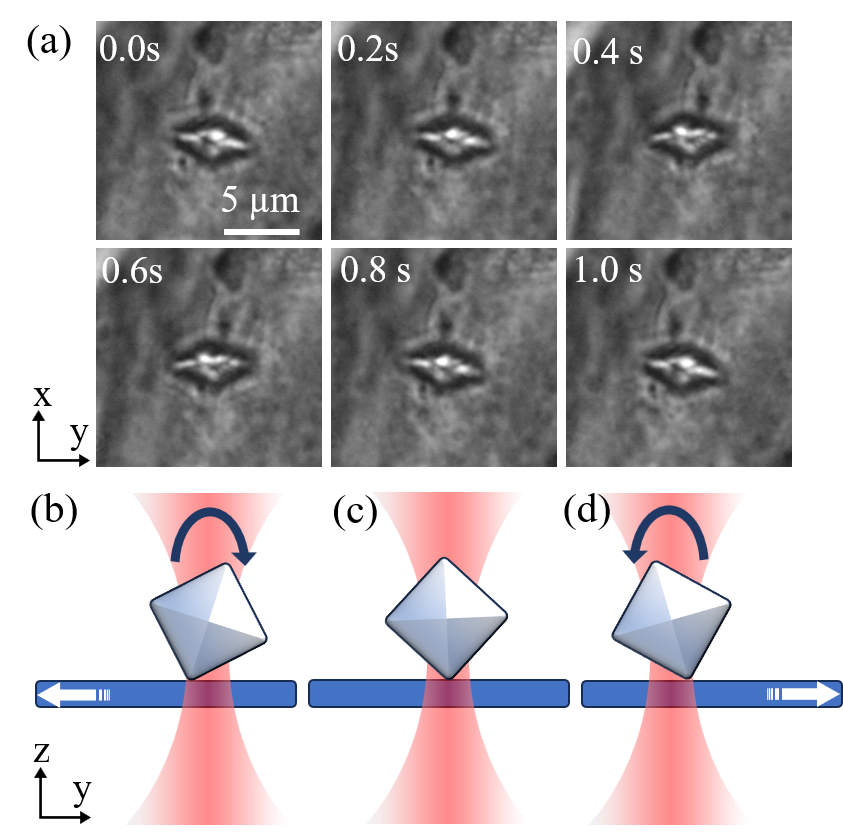}
\caption{(a) Roll rotation of CaOx particle inside plant cell using stage oscillation. The particle rotates clockwise when the stage moves left from equilibrium (b) and counterclockwise when the stage moves right (d). The equilibrium position, with no stage motion, is shown in (c)}
\label{video}
\end{figure}

\begin{figure}[!ht]
\centering\includegraphics[width= \linewidth]{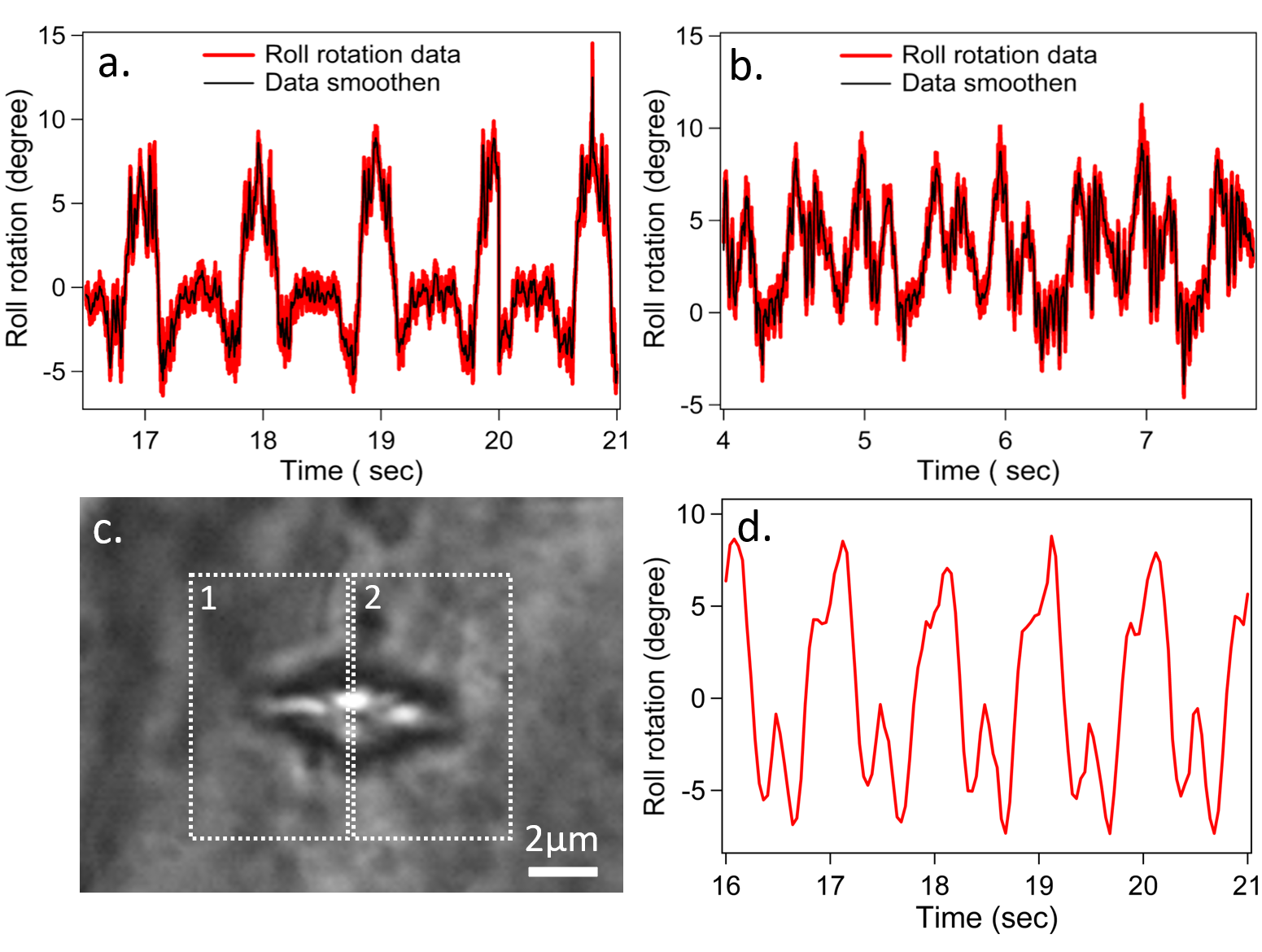}
\caption{Time series for the roll rotation of CaOx particle (a)  when the orientation of the crystal is mainly about $\phi$ = 45° and the crystal tilts small amounts about this orientation (b) When the general orientation of the crystal has been turned slightly off from the $\phi$= 45°, and then turned small extents about the mean (c) A typical image of a particle placed on the plasma membrane surface, and subsequent estimation of roll rotation using an alternative strategy of image processing. (d) The obtained estimates of the roll angle from the alternative strategy. The time series matches well with estimates from photonic force microscopy. The corresponding theoretical plot is shown in FIG. \ref{fig:FOUR LOBE}. } 
\label{time_series}
\end{figure}

The particle optically trapped inside the leaf cell was first placed close to the plasma membrane surface, and then the stage moved parallel to that surface. This then turns the optically confined particle due to the frictional properties of the membrane surface. We show the images of such particle turning while being optically confined and placed on the bottom surface of the leaf cell membrane in FIG. \ref{video}. The typical time series of such rotation is shown in FIG. \ref{time_series}. The edge of the crystal can be assumed to be hemispherical, with a diameter of approximately 500 nm. Hence the contact diameter with the surface might be about 200 nm, making it suitable for nano-tribological investigations. 

The time series and the signal shown in FIG. \ref{time_series}(a)  matches well with the theoretically predicted response function which shall be presented in a later section. We also show another corresponding rotation signature extracted from the video images. We divide the image of the particle into two halves, which are labeled  as 1 and 2 (FIG. \ref{time_series}(c)), and then calculate the total grayscale value of all the pixels within them. We then subtract the total grayscale intensity of box 1 from box 2, which indicates the extent of rotation of the particle\cite{ghosh2025determination}. This time series signal, shown in FIG. \ref{time_series}(d), matches well with that extracted from photonic force microscopy, indicated in FIG. \ref{time_series}(a). The rotation due to the stage motion in this figure possibly does not comprise the full $\pm$20° to complete the linear range shown in FIG. \ref{fig:FOUR LOBE}. 

\begin{figure}[!ht]
\centering\includegraphics[width=\linewidth]{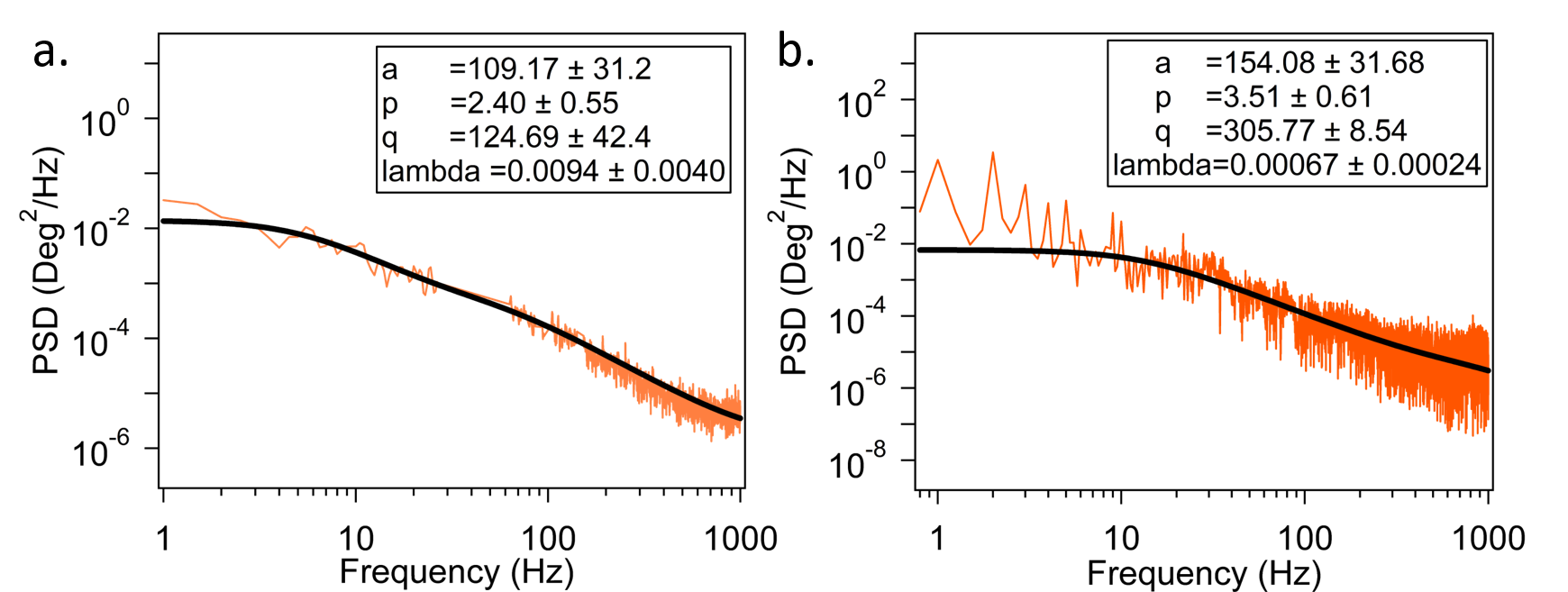}
\caption{(a)  PSD obtained  for a CaOx crystal in the bulk (b) PSD of the same CaOx crystal near the cell plasma membrane} 
\label{psd}
\end{figure}

In the time series, it can be noticed that the transfer function is close to the middle region of the curve, where the crystal is supposed to orient at $\phi$= 45°, is linear. Hence, any small displacements around the mean at this location of the roll angle are transferred into the detected signal. We can also note that, as the mean orientation of the crystal changes from $\phi$ = 45° to other values, the time series gets skewed to one side (FIG. \ref{time_series}(b)), proving the non-linearities of the response function indicated in the theory section. We can estimate the power spectral density (PSD) of the roll angle well around the mean for this configuration and is shown in FIG. \ref{psd}.

These power spectral density datasets (PSD) have been fitted using the Generalised Maxwell model\cite{ferrer2021fluid,raikher2013brownian,grimm2011brownian}. The general functional form for the PSD is given in Eq. (\ref{psda}). There are 2 PSDs shown in FIG. \ref{psd}. In (a), we show the response in  the bulk, whereas, the response in (b) shows the PSD when the crystal has been placed closer to the leaf cell membrane surface, with the mean orientation at $\phi$ = 45\textdegree. 

The frequency-dependent viscosity for an incompressible, low Reynolds number, linear viscoelastic fluid composed of a solvent and a dissolved polymer solute, is obtained from the Stokes–Oldroyd-B model of linear viscoelasticity\cite{roy2023comparison,paul2018free}.

\begin{equation}
\mu(\omega)= \mu_s + \frac{\mu_p}{ 1 -i \omega \lambda}
\label{model}
\end{equation}

Here, the $\mu_s$ denotes the zero-frequency solvent viscosity, $\mu_p$ represents zero-frequency polymer viscosity, and $\lambda$ is the polymer relaxation time. Then the roll rotation PSD of the Brownian motion inside such a fluid is given by the expression

\begin{equation}
\mathrm{PSD}(\omega) = 
\frac{2 k_B T  \left[ \left( 1 + \frac{\mu_p}{\mu_s} \right) \lambda^2 + \omega^2 \right]}
{\gamma\left[ \left( \frac{\kappa}{\gamma \lambda }- \omega^2 \right)^2 
+ \omega^2 \left(\frac{\kappa}{\gamma} + \frac{1}{\lambda} \left( 1 + \frac{\mu_p}{\mu_s} \right) \right)^2 \right]}
\label{psda}
\end{equation}

This equation can also be written as

\begin{equation}
\mathrm{PSD}(\omega) = 
\frac{A  \left[ p \lambda^2 + \omega^2 \right]}
{\left[ \left( \frac{q}{ \lambda }- \omega^2 \right)^2 
+ \omega^2 \left(q + \frac{p}{\lambda} \right)^2 \right]}
\label{psd1}
\end{equation}
\begin{equation}
p = \left( 1 + \frac{\mu_p}{\mu_s} \right), \quad 
q = \frac{\kappa}{\gamma}, \quad 
A = \frac{2 k_B T}{\gamma},
\label{psd coefficients}
\end{equation}

where, $\kappa$ represents the angular trap stiffness, $\gamma$ is the rotational drag coefficient for the solvent. The coefficient A signifies the amplitude in terms of Volt$^2$/ Hz. By fitting with the Eq.(\ref{psda}), the values of  $  p$, $q$, $A$ and $\lambda$ can be obtained.The calibration factor $\beta$ in (rad/Volt)$^2$ is quite similar to the conventional calibration factor for normal media, given as
\begin{equation}
\beta = \frac{2 k_B T}{A \gamma}
\label{drag}
\end{equation}
The experimentally acquired PSD signals are multiplied by $\beta$, then converted to units of Degree$^2$/Hz. Similarly, time series signals are multiplied by $\sqrt{\beta}$ and then converted to units of Degrees.
The particle is assumed to be an oblate spheroid to calculate the drag coefficient for roll rotation. The semi-major axis (a) of the crystal trapped in the experiment is 2.5 $\mu$m, and the semi-minor axis (c) is 0.5 $\mu$m. According to Perrin's drag formulas\cite{perrin1934mouvement}, the multiplicative factor $F_{ax}$ for rotation along the semi-minor axis is 2.67, and the effective radius of the sphere, $a_0$, is 1.46 $\mu$m. So, the drag coefficient for roll rotation at bulk is 
\begin{equation}
\gamma_0 = F_{ax}\, 8\pi\mu_s\, a_0^{\,3}, 
\qquad 
a_0^{\,3} = a^2 c
\label{drag_shape}
\end{equation}
The Stokes drag force acting on a rotating microparticle is increased by the presence of a cell membrane by a factor given by Faxen's correction\cite{leach2009comparison}. When the rotation axis is perpendicular to the surface normal, the roll rotational drag coefficient can be written as  
\begin{equation}
\gamma = \frac{\gamma_0}{1 - \frac{5}{16}\left(\frac{a}{s}\right)^3 + \frac{15}{256}\left(\frac{a}{s}\right)^6}
\ 
\label{faxen correction}
\end{equation}
For a particle-membrane separation of 100 nm, the drag coefficient $\gamma$ is estimated to be 1.3 times $\gamma_0$.
A typical fit to the PSD for roll rotational motion of the crystal away from the surface of a leaf cell membrane is shown in FIG. \ref{psd}(a). Similarly, a typical fit to a PSD of roll rotational motion for such a crystal touching the inner surface of the leaf cell membrane is shown in FIG. \ref{psd}(b).The angular trap stiffness($\kappa$) can be obtained from the fitting parameter q and Eq.(\ref{psd coefficients}). The value of $\kappa$ is $42.5$ pN.$\mu$m/rad and the torque is $7.4$ pN.$\mu$m when the crystal is nearly touching the inner surface of the membrane. Since the crystal's radius is 2.5$\mu$m, the effective force acting on the crystal is 18.5 pN, for which the crystal is turning by about 10 degrees. However, we also see from the power spectral density that the system can sense about 0.1 degrees at a bandwidth of 1 Hz, such that the force it senses is about 200 fN. Thus the frictional force measured is 18.5 $\pm$ 0.2 pN.

The parameters extracted from these PSDs can be used to ascertain the viscoelastic properties of the leaf cell interior. Using the fits, we find that the values of the p and $\lambda$,  when the crystal is in the bulk of the cell as 2.40 and 0.009 $\pm$ 0.004 sec respectively. The same coefficients when the crystal is nearly touching the inner surface of the cell membrane are 3.51 and 0.0007 $\pm$ 0.0002 sec respectively. The presence of the cell membrane surface in proximity to the crystal can be interpreted better by adding an extra Maxwell element to the viscosity expression, which then becomes,

\begin{equation}
\mu(\omega)= \mu_s + \frac{\mu_p}{ 1 -i \omega \lambda} +\frac{\mu_{p1}}{ 1 -i \omega \lambda_1}
\label{model1}
\end{equation}

The $\lambda$ is one order of magnitude larger than $\lambda_1$ which is the relaxation near the membrane. We assume that the leaf friction term is much larger than the second term, and hence one may assume that Eq. \ref{model1}
becomes

\begin{equation}
\mu(\omega)= \mu_s +\frac{\mu_{p1}}{ 1 -i \omega \lambda_1}
\label{model2}
\end{equation}

Then the corresponding expression for p becomes $ 1 + \frac{\mu_{p1}}{\mu_s }$. Since the value of p when the crystal was in the bulk was 2.4, the viscosity of the interior of the leaf cell is estimated to be about 1.4 $\mu_s$, where the $\mu_s$ is assumed to be that of water. Further, since the value of p when the crystal was touching the cell membrane surface is 3.51, the viscosity of the fluid near the membrane surface is obtained as $\mu_{p1}$ =2.51 ($\mu_s$). This value is larger than the value of the bulk viscosity and hence the corresponding term in Eq. \ref{model1}, would dominate in the expression for viscosity, when the crystal is touching the cell membrane. Moreover, we also find that the PSD on the surface has a lower time constant $\lambda$ than that in bulk, which can only be possible when the term due to the surface friction dominates over the second term in Eq. (\ref{model1})

Thus, the final expression for the frequency-dependent viscosity when the crystal is touching the leaf cell membrane surface can be obtained as,

\begin{equation}
\mu(\omega)= \mu_s + \frac{1.4 \mu_s}{ 1 -i 0.0094 \omega } +\frac{2.51 \mu_s}{ 1 - i  0.00067 \omega }
\label{result1}
\end{equation}
\begin{equation}
\mu(\omega)\approx \mu_s +\frac{2.51 \mu_s}{ 1 - i  0.00067 \omega }
\label{result2}
\end{equation}
\begin{equation*}
G'= \omega\    Im(\mu)
\end{equation*}
\begin{equation}
G''= \omega\    Re(\mu)
\label{result3}
\end{equation}

\begin{figure}[!ht]
\centering\includegraphics[width=0.8\linewidth]{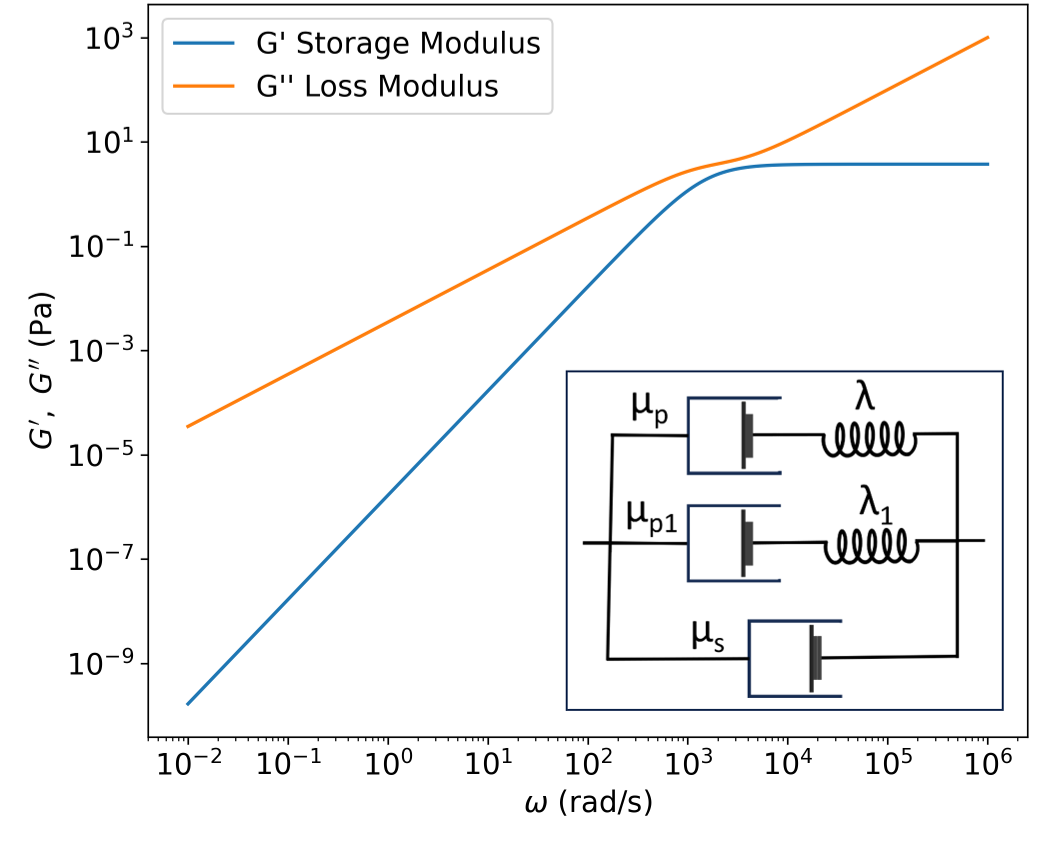}
\caption{The storage modulus, G', and loss modulus, G'', curves for the CaOx crystal rotation inside the leaf cell obtained using the model shown in the inset. The inset shows a model with an extension to the generalised Maxwell model to include the effect of the surface friction.}
\label{fig:Gpp}
\end{figure}

The coefficient of the last term in Eq.(\ref{result1}) is a factor of 2 higher than the second term, and hence the second term may be neglected to obtain the effect of the membrane surface on the crystal.
We can derive the storage modulus (G') and loss modulus (G") from Eq. (\ref{result3}).The plot of the G' and G'' as a function of frequency are shown in FIG. \ref{fig:Gpp}. 

As can be observed in FIG. \ref{fig:Gpp}, the viscous part dominates the elastic part at all frequencies. This is consistent with the observations that, as the particle does not stick to the surface, it does not have an elastic contribution. These plants are semi-succulent, and the leaves store increased amounts of water, which may explain the consistently higher loss modulus (G'') observed over the measured frequency range. Moreover, the surface may have fluids that may prevent easy sticking of the particle to the surface. Further, the absolute values of the G' and G'' are consistent with similar reports in the literature \cite{sharma2008microrheology}. The inset provides an overview of the model used. 

The plant cell wall is made up of cellulose fibers along with other polysaccharides, such as pectin, hemicellulose etc. along with other sensory proteins. It seems that the net effect of such a surface is that in the time scales that we performed the experiment, the particles do not adhere to the surface, but rather experience a viscous frictional force. This kind of observation is useful to address many important questions, like morphogenesis and the variation of water inside the cell, which would affect the frictional properties.

\section{\label{sec:level6}CONCLUSIONS:}

Thus, to conclude, we use calcium oxalate particles naturally occurring inside \textit{Tradescantia zebrina} leaf cells to understand the cell fluid properties \textit{in vivo} without damaging the cell wall. These particles trap in a particular configuration inside the optical tweezers and then we study the out-of-plane (or roll) rotational degree of freedom to estimate the frictional properties of the inner surface of the cell. This strategy, to use roll rotation to study frictional properties, has never been explored before in plant cell investigations. This paves way for further studies on the plant cells under different conditions, such as, water stress, the effect of application of chemicals, various plant pathogenic conditions and even growth or development of cells.

\section{\label{sec:level7}ACKNOWLEDGEMENTS:}

We thank the Indian Institute of Technology, Madras, India, for their seed and initiation grants to Basudev Roy. This work was also supported by the DBT/Wellcome Trust India Alliance Fellowship IA/I/20/1/504900 awarded to Basudev Roy.

\section{Images and Graphics}
 All the images and the graphics used in this manuscript are freshly made and not published anywhere else.

 \section{\label{sec:level8}Materials and Methods:}

\subsection{Experimental Setup}
\begin{figure}[!ht]
\centering\includegraphics[width=.9 \linewidth]{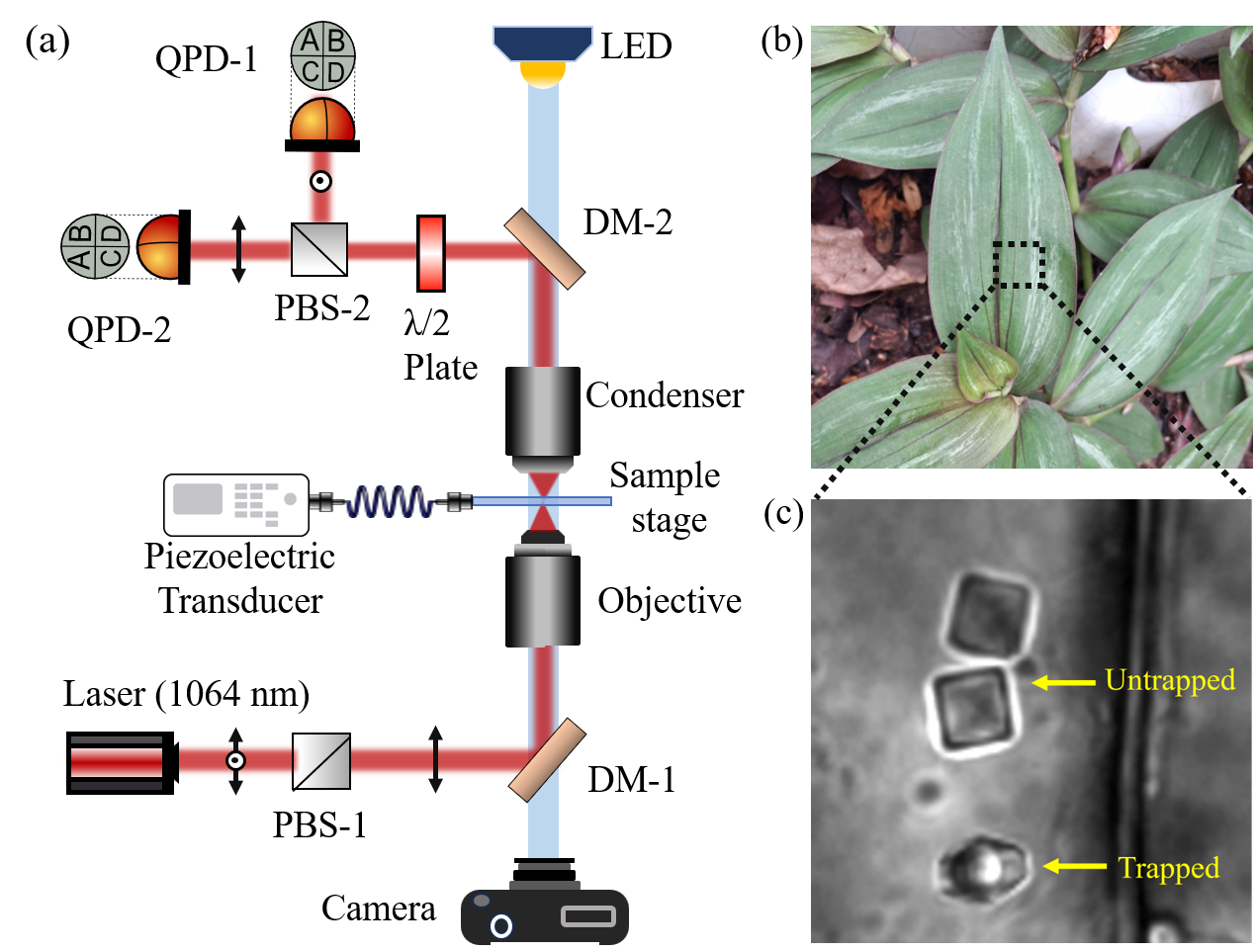}
\caption{(a) Schematic diagram of the Optical tweezer setup.(b) Image of the \textit{Tradescantia zebrina} leaf used in the experiments (c)Calcium oxalate crystals inside an epidermal cell of \textit{Tradescantia zebrina} leaf.
The crystals exhibit a square facet when observed in the untrapped state. Upon optical trapping, the crystals reorient and reveal the hexagonal facet }
\label{fig:ot schematic}
\end{figure}

The optical tweezers setup used in the experiment is a modified OTKB/M kit (Thorlabs, USA) operated in an inverted configuration, as shown in FIG. \ref{fig:ot schematic}. The setup integrates a 100x, 1.3 Numerical Aperture (N.A.) Olympus oil-immersion objective and an E Plan 10x, 0.25 N.A. A Nikon air-immersion condenser lens forms the microscopy unit. A diode laser of 1064 nm wavelength (Lasever  Systems) is employed as a trapping laser. The laser is directed into the sample chamber using a polarizing beam splitter and a dichroic mirror, and then tightly focused on the sample by the objective. Forward scattered light from the sample plane is collected by a condenser and directed through another dichroic mirror, polarizing beam splitter to a couple of quadrant photo-diodes (QPDs). The QPD-1 detects the cross-polarized forward-scattered light, and QPD-2 detects the same polarization light as the incident polarization. QPD-1 gives us information about the rotational motion of the trapped particle, whereas QPD-2 gives details about the translational motion.

 A white LED illuminates the chamber from above, with light collected by a CMOS camera (Thorlabs) through DM-1 for imaging. This system enables precise trapping and manipulation of particles, including the detection of rotational degrees of freedom such as pitch and roll, which is critical for applications in microrheology, cell membrane studies, active matter systems, etc.

\subsection{Experimental Method}

First, we prepare a leaf epidermal peel by carefully tearing the \textit{Tradescantia zebrina} leaf from the edge. After that, we place a thin layer of the peeled epidermal cells on a coverslip. We illuminate the 1064 nm wavelength with 50 mW power at the sample plane to trap a single calcium oxalate particle inside an epidermal cell of a leaf. We modulate the sample stage at 1 Hz using a a function generator and piezo-electric stage. Since the CaOx particle is trapped and is in contact with the cell plasma membrane, the particle starts rotating in a rolling fashion. The prismatic shape of the particle helps us to detect the rotation angle. We obtain translational and rotational signals from the forward scattered light and microscopic images from the back-scattered light. The experiments were carried out five times to confirm the observation and the results.



\bibliographystyle{unsrt}
\bibliography{apssamp}

\end{document}